\documentclass[twocolumn]{aastex63}

\usepackage{gensymb}
\usepackage{xspace}
\usepackage{amsmath}

\newcommand{\secref}[1]{Section \ref{#1}}
\newcommand{\figref}[2]{Figure \ref{#1}{#2}\xspace}
\newcommand{\unit}[1]{\,\ensuremath{\mathrm{#1}}}

\renewcommand{\vec}[1]{\ensuremath{\mathbf{#1}}}

\newcommand{\eg}{\textsl{e.g.}\xspace}
\newcommand{\kmps}{\unit{km\cdot s^{-1}}}

\newcommand{\dquote}[1]{``{#1}''\xspace}

\newcommand{\AR}{AR 12838\xspace}

\newcommand{\rev}[1]{{\bf#1}}

\received{ }
\revised{ }
\accepted{ }
\submitjournal{ApJL}

\shorttitle{light bridge}
\shortauthors{Feng et. al.}

\begin{document}

\title{Magnetic connectivity between the light bridge and penumbra in a sunspot}

\correspondingauthor{Ding Yuan}
\email{yuanding@hit.edu.cn}

\author[0000-0003-4709-7818]{Song Feng}
\affiliation{Yunnan Key Laboratory of Computer Technology Application, Faculty of Information Engineering and Automation, \\ Kunming University of Science and Technology, Kunming 650500, China}
\affiliation{Yunnan Astronomical Observatory, Chinese Academy of Sciences, PO Box 110, Kunming 650011, China}

\author{Yuhu Miao}
\affiliation{Institute of Space Science and Applied Technology,
	Harbin Institute of Technology, Shenzhen,
	Guangdong 518055, China}

\author[0000-0002-9514-6402]{Ding Yuan}
\affiliation{Institute of Space Science and Applied Technology,
	Harbin Institute of Technology, Shenzhen,
	Guangdong 518055, China}

\author{Zhongquan Qu}
\affiliation{Yunnan Astronomical Observatory, Chinese Academy of Sciences, \\
	PO Box 110, Kunming 650011, China}

\author{Valery M. Nakariakov}
\affiliation{Centre for Fusion, Space and Astrophysics, Department of Physics, University of Warwick, CV4 7AL, UK}
\affiliation{St Petersburg Branch, Special Astrophysical Observatory, Russian Academy of Sciences, St Petersburg, 196140, Russia}



\begin{abstract}
A light bridge is a prominent structure commonly observed within a sunspot. Its presence usually triggers a wealth of dynamics in a sunspot, and has a lasting impact on sunspot evolution. However, the fundamental structure of light bridges is still not well understood. In this study, we used the high-resolution spectropolarimetry data obtained by the Solar Optical Telescope onboard the Hinode satellite to analyze the magnetic and thermal structure of a light bridge at \AR. We also combined the high-cadence $1700\unit{\AA}$ channel data provided by the Atmospheric Imaging Assembly onboard the Solar Dynamic Observatory to study the dynamics on this bridge. We found that a pair of blue and red Doppler shift patches at two ends of this bridge, this pattern appears to be the convective motion directed by the horizontal component of the magnetic field aligned with the spine of this bridge.  Paired upward and downward motions implies that the light bridge could have a \rev{two-legged} or undulate magnetic field.  Significant four minute oscillations in the emission intensity of the $1700\unit{\AA}$ bandpass were detected at two ends, which had overlap with the paired blue and red shift patches. The oscillatory signals at the light bridge and the penumbra were highly correlated with each other. Although they are separated in space at the photosphere, the periodicity seems to have a common origin from the underneath. Therefore, we infer that the light bridge and penumbra could share a common magnetic source and become fragmented at the photosphere by magneto-convection.
\end{abstract}


\section{Introduction} 
\label{sec:intro}

A light bridge is an elongated bright structure that crosses a sunspot umbra or deeply penetrates into it \citep{Muller1979}. This type of structure is usually found in a nascent or decaying spot, or when two spots merge into one or the inverse process occurs. At these stages, magneto-convection causes significant flux separation in deeper layers of a spot or pore, and this phenomenon could be accompanied by the formation of light bridges and umbral dots at the photosphere \citep{Bharti2007a,Rempel2011,Toriumi2015a}.  Light bridges could affect the evolution of sunspots, \eg, split or merge, they consequently determine the level of solar activities indirectly. 

According to \citet{Sobotka1997}, a light bridge is classified either as a granular bridge if it exhibits photospheric dynamics, or a filamentary bridge if it is an intrusion of penumbra. Further classification is termed either as a strong light bridge if it separates the umbra, or otherwise a faint light bridge if it simply protrudes into the umbra. In the following text, we use \dquote{spine} to refer the elongated central column of a light bridge. Two \dquote{ends} are a bridge's shorter edges, which normally have a length of a few arcsec, and two \dquote{flanks} are the longer edges, which usually has a length scale comparable with the umbra.

A light bridge is believed to be hotter than the umbra, as convection (granulations) are partially or fully restored therein \citep{vanderVoort2010}. Convection could act as an effective mechanism to supply hotter gas to the light bridge. Hot and dense plasma is pushed upwards at the spine of a bridge and moves downwards at its two flanks \citep{Lagg2014}. Strong downflow would drag the magnetic field lines and could even lead to polarity reversal in extreme cases \citep{Bharti2007b}. 

The magnetic field of a light bridge is more inclined than the vertical umbral field; a magnetic canopy model is proposed for light bridge \citep{Jurcak2006}. Strong discontinuity is found at the interface between a light bridge and the umbra, therefore, strong electric current density could be detected at the edge of a bridge \citep{Shimizu2011,Toriumi2015a,Toriumi2015b}. 

Many dynamic activities are detected at light bridges, as light bridge has a magnetic and thermal structure in contrast with that of the umbra. \citet{Louis2014} found that dynamic jets usually originate from one flank of a bridge and propagate along the open magnetic field lines. According to the asymmetry in jet excitation, \citet{Yuan2016lb} proposed a three dimensional model for a light bridge: the magnetic field at one flank is mostly aligned with umbral field, whereas at the other flank the two magnetic fields are more close to anti-parallel. In this scenario, magnetic reconnections, which are believed to be the driver of jets \citep{Moore2010}, could only be triggered at one flank. 
  
\citet{Robustini2016} and \citet{Tian2018} detected inverted Y-shaped jets above sunspot light bridges, this morphology supports the model of flux emergence in the vertical umbral magnetic field, the collimated jets and surges are propelled by the reconnections between anti-parallel magnetic field lines. This kind of activities are also reported by a number of studies \citep{Bharti2017,Bai2019}. 

A sunspot is usually a host of waves and oscillations. In the umbra, the oscillation period is about three minutes, whereas in the penumbra, a dominant oscillation could be found at five minute bandpass \citep{Khomenko2015}. Above a light bridge, five minute oscillations similar to penumbra oscillations are usually detected, however, its origin is still not revealed yet \citep{Yuan2014,Yuan2016lb,Yang2015}.

In this study, a tiny light bridge is formed in the center of the umbra, the ambient magnetic field is very simple, this situation hence allows a decent investigation of light bridge properties with negligible contamination of other structures. In \secref{sec:obs}, we presents data reduction and analysis. Then we cover the structure and dynamics on this light bridge in \secref{sec:results}. In \secref{sec:con}, we discuss a possible structure of light bridge with the knowledge obtained in this study.

\section{Data reduction and analysis}  
\label{sec:obs}

\begin{figure}[ht]
\centering
\includegraphics[width=0.5\textwidth]{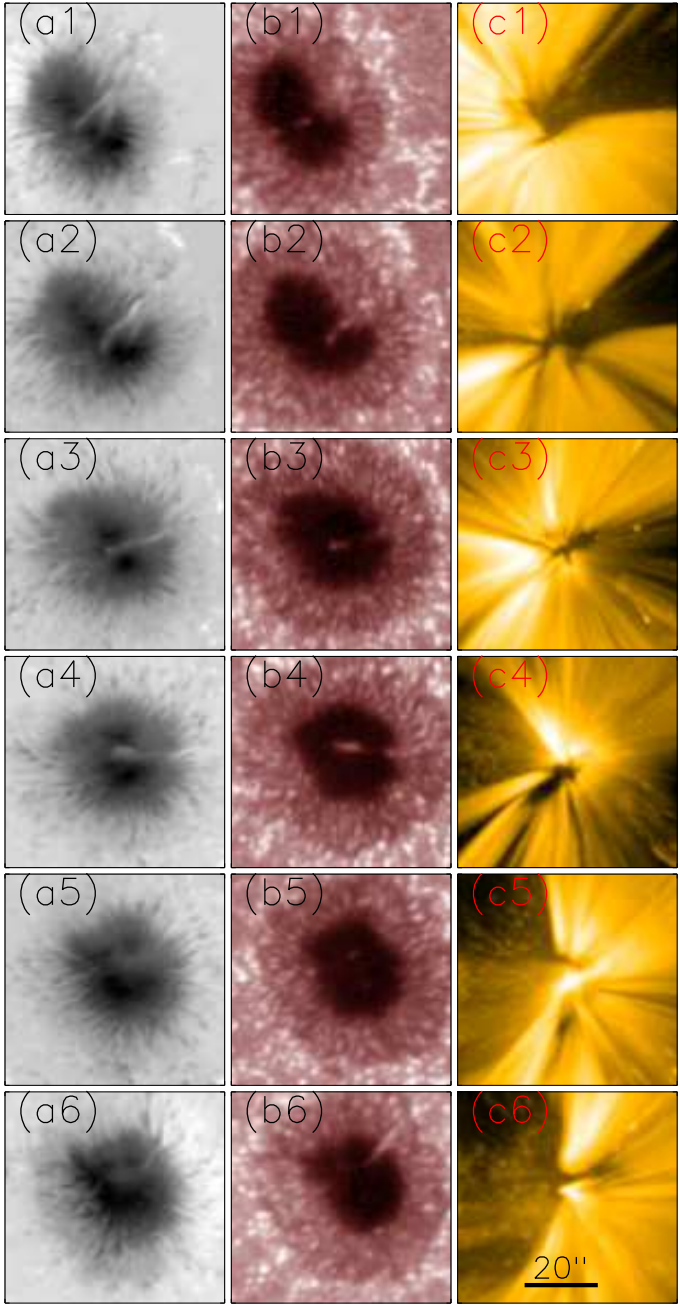}
\caption{Evolution of the sunspot and diffuse coronal loops associated with active region  12738. The left (a), middle (b), and right (c) columns draw the images of HMI LOS magnetogram, AIA 1700 \AA{}, and AIA 171 \AA{} channels, respectively. Number 1-6 denotes the observations at about 10:00 UT on consecutive dates from $10^\mathrm{th}$ to $15^\mathrm{th}$ April 2019. \label{fig:fov_evol}}
\end{figure}

On 10 and 11 April 2019, a penumbral filament protruded into the umbra of \AR and formed a faint light bridge (see \figref{fig:fov_evol}). This light bridge detached from the penumbral filament and evolved into a compact and isolated structure on 12 and 13 April. From 14 April and thereafter, the light bridge became more diffuse and eventually decayed off. This light bridge was clearly visible in the $1700\unit{\AA}$, $1600\unit{\AA}$, and visible light ($4500\unit{\AA}$) channels of the Atmospheric Imaging Assembly \citep[AIA,][]{AIA2012} onboard the Solar Dynamics Observatory (SDO) from 10 to 15 April 2019.

SDO/AIA observes the full solar disk continuously with nine filters in ultra violet (UV) and extreme UV (EUV) bandpasses. One AIA pixel correspond to an angular width of $0.6\unit{arcsec}$ or a distance of about $435\unit{km}$ on the sun. The UV channels take an image of the Sun every $24\unit{s}$, whereas the EUV images have a cadence of about $12\unit{s}$. 

The Helioseismic and Magnetic Imager (HMI) onboard SDO observes the full solar disk at $6173\unit{\AA{}}$, one pixel corresponds to $0.5\unit{arcsec}$. The line-of-sight magnetogram has a cadence of about $45\unit{s}$.

The AIA UV and EUV data were calibrated with the standard procedure provided by the SolarSoft\footnote{\url{http://www.lmsal.com/solarsoft/ssw_install.html}}. The digital offset of the cameras, CCD read-out noise and dark current were removed from the data, then each image was corrected with a flat-field and was normalized with its exposure time.

The spectro-polarimeter (SP) of the Solar Optical Telescope (SOT) onboard the Hinode mission measured high-precision Stokes polarimetric line profiles of the Fe I $6301.5\unit{\AA{}}$ and $6302.5\unit{\AA{}}$ spectral lines. The fast-mapping mode was used to scan the entire sunspot and its surrounding area. Each pixel represented an angular width of about $0.32\arcsec$; the spectral sampling was about $21.549\unit{m\AA{}}$ per pixel. Two consecutive slit scan had a step size of about $0.30\arcsec$. 

We obtained SOT/SP level-2 data from Community Spectro-polarimetric Analysis Center (CSAC\footnote{\url{https://csac.hao.ucar.edu/sp_data.php}}). This dataset has undergone the standard CCD image calibration. The Stokes parameters were used to derive the magnetic field vector and flow field by assuming the Milne-Eddington Atmosphere model \citep[see review by][]{Iniesta2016}. The $180\degree$ ambiguity of the azimuth angle of magnetic field was resolved with the AZAM utility \citep{Lites1995}. The SP images were aligned with the AIA UV images by matching the center of the light bridge. The vertical electric current density ($j_z$) were computed by using the inverted vector magnetic field, $j_z=\mu_0^{-1}(\nabla\times\vec{B})_z$, where $\mu_0$ stands for the magnetic permeability in free space. The Doppler shift was obtained by the inversion of Stokes vector.  

\begin{figure*}[ht]
\centering
\includegraphics[width=0.7\textwidth]{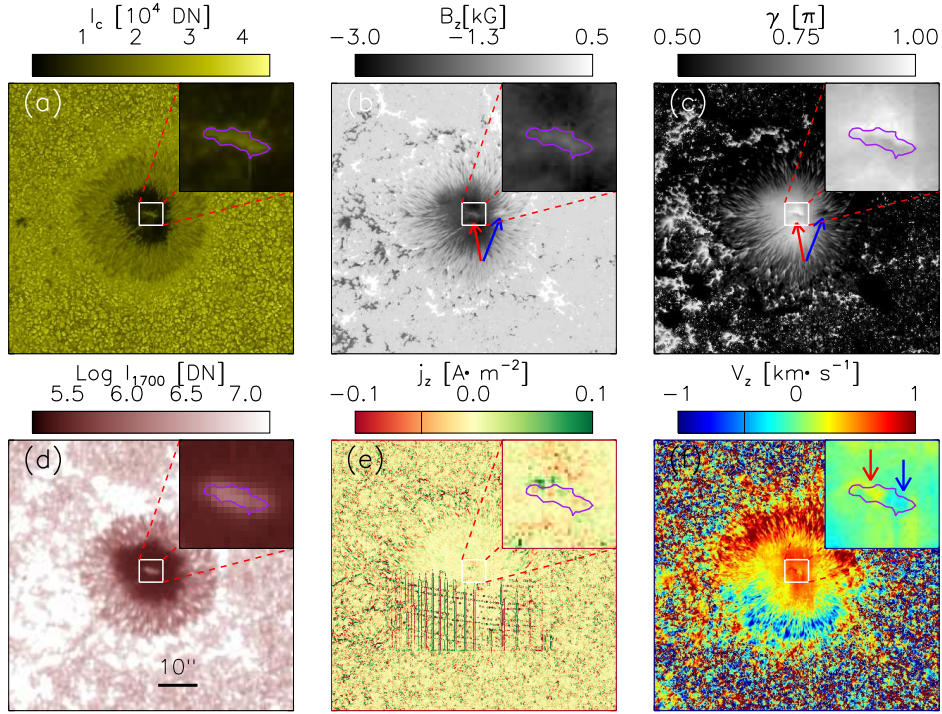}
\caption{Maps of the SOT/SP continuum intensity at $\lambda6301.5\unit{\AA{}}$ (a), the LOS magnetic field component (b), the inclination angle of magnetic field vector (c), the electric current density (e), the Doppler shift velocity (f) and the emission intensity of the AIA $1700\unit{\AA{}}$ channel (d). The top-right window of each panel highlights the zoomed-in view of the light bridge. A pair of red and blue arrows highlight the light bridge and another region with mixed polarity at the penumbra. In zoom-in window of (f), the local background velocity has been removed to highlight the Doppler shift above light bridge. \label{fig:fov}}
\end{figure*}

\begin{figure*}[ht]
\centering
\includegraphics[width=0.7\textwidth]{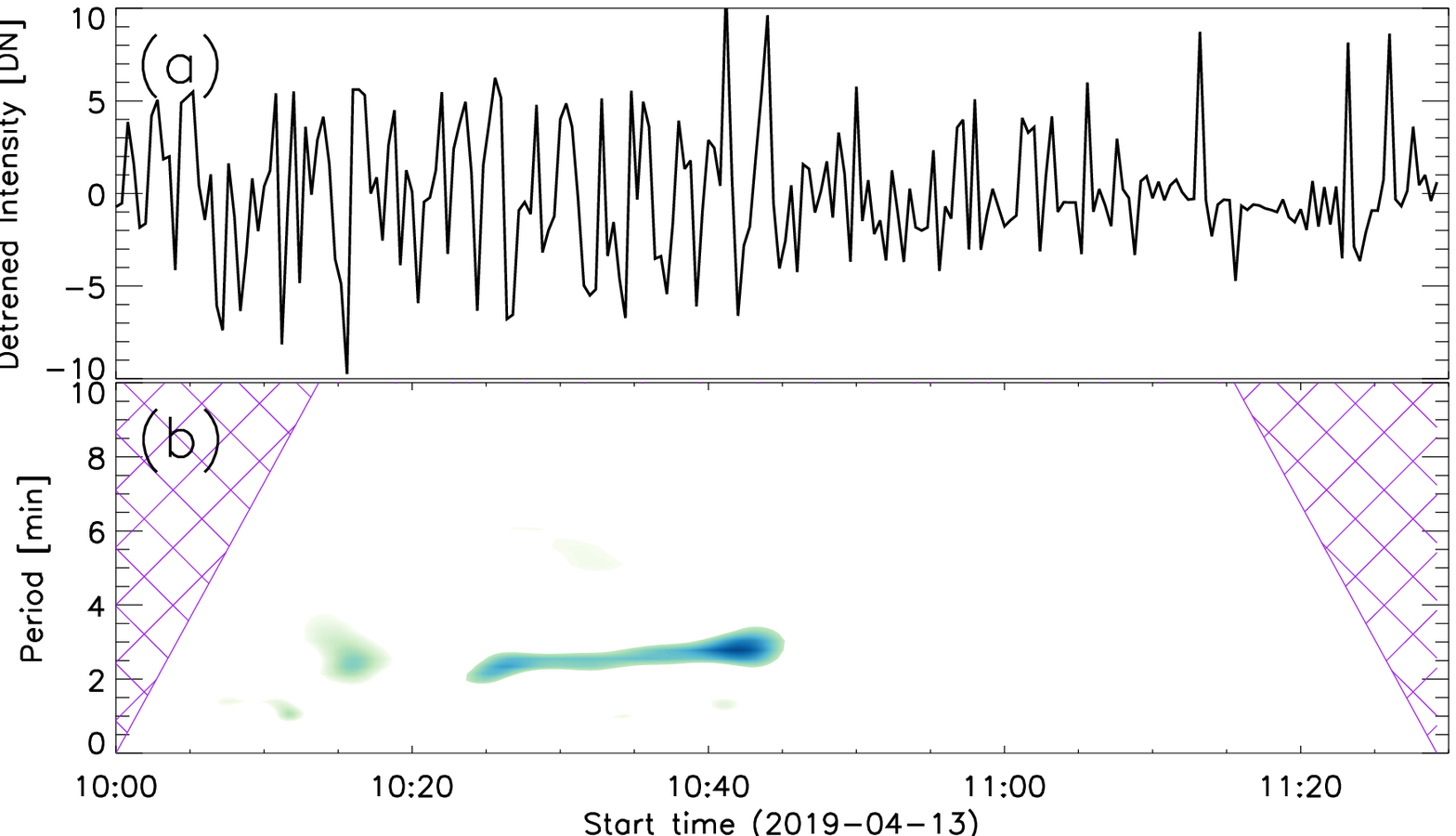} 
\includegraphics[width=0.7\textwidth]{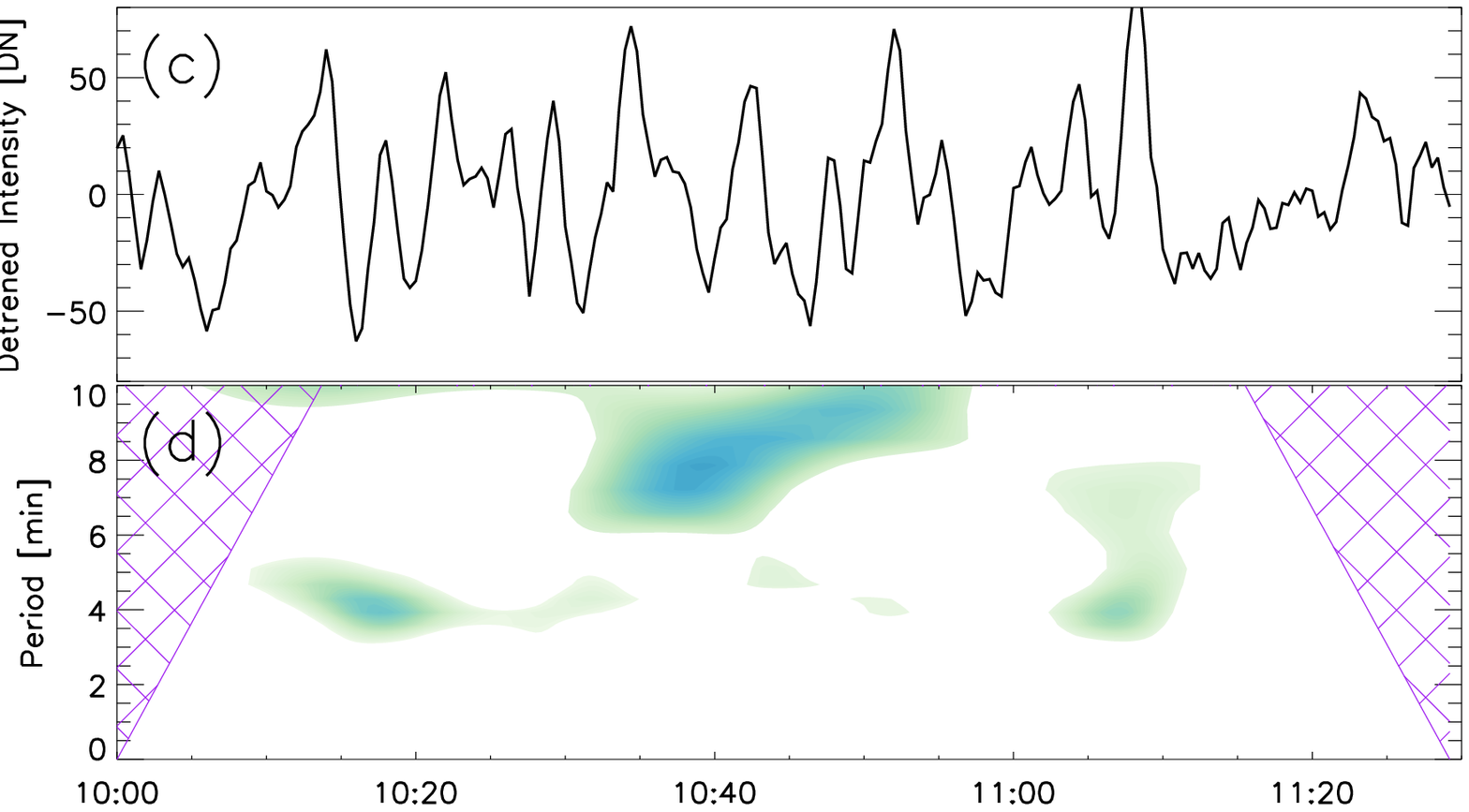}
\caption{(a) Emission intensity of the AIA $1700\unit{\AA}$ channel at a pixel on the umbra, ten minute running average were removed. (b) Wavelet power spectrum of the de-trended signal. The reliable regions subjective to zero-padding effect is cross-hatched. (c) and (d) Same analysis done to a pixel on the light bridge. \label{fig:wavelet}}
\end{figure*}

We extracted the emission intensity variations of the $1700\unit{\AA}$ channel from a pixel in the umbra and another one at the light bridge (\figref{fig:fov}{d}). These two time series were detrended by removing the ten minutes moving average; the wavelet spectra were calculated with the Morlet mother function \citep{Torrence1998}. The detrended emission intensity variations and wavelet spectra are plotted in \figref{fig:wavelet}{} for both cases of light bridge and umbra. We see that an oscillatory signal with a period of about two minutes was detected in the umbra (\figref{fig:wavelet}{b}), whereas the light bridge exhibited a four minute oscillation (\figref{fig:wavelet}{d}). 

\begin{figure*}[ht]
\centering
\includegraphics[width=0.7\textwidth]{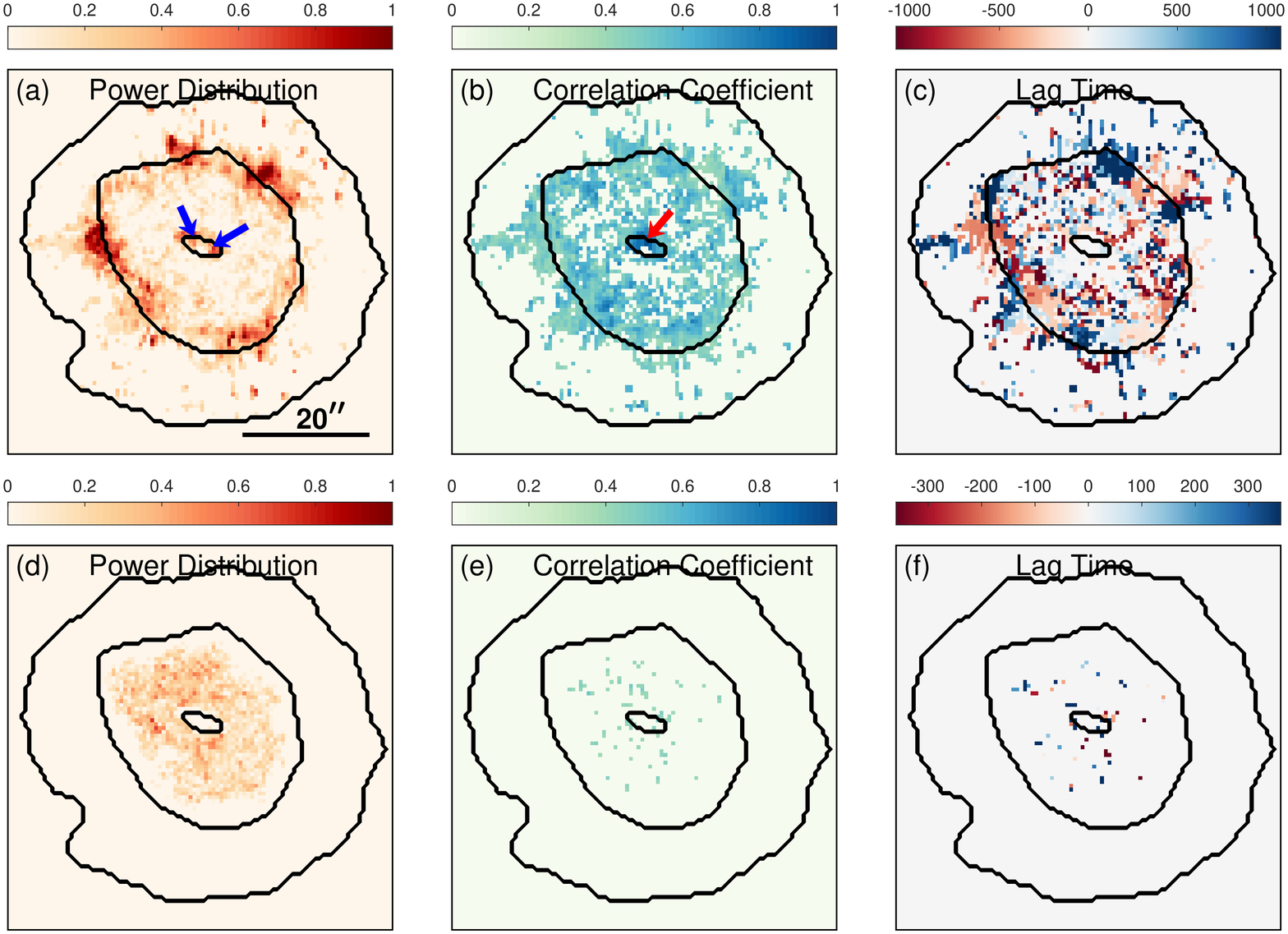}
\caption{(a) Fourier power averaged over three to five minute period range in AIA $1700\unit{\AA}$ channel. (b) and (c) are the maximum cross-correlation coefficient and lag time for the time series filtered at the same bandpass. (d)-(f) are the counter part for one to three minute period range. \label{fig:powermap}}
\end{figure*}

To study the spatial distribution of four minute oscillations, we performed Fourier transform to the detrended emission intensity of every pixel in the AIA $1700\unit{\AA}$ and averaged the Fourier power among the spectral components between three and five minutes. \figref{fig:powermap}{a} plots the narrowband Fourier power with four minute spectral peak above $3\sigma$ noise level. The noise level was estimated by assuming the white noise in the times series as formulated in \citet{Torrence1998}.

We also analyzed the correlation between the four minute oscillations at the light bridge and those at the penumbra. For every pixel, the detrended emission intensity was transformed into the Fourier space, then a narrow band spectrum was selected by multiplying a uniform window between three to five minute. Thereafter, a filtered signal in time domain was formed by doing inverse Fourier transform. A reference signal was estimated by averaging the signals within the light bridge (within the island as enclosed by the inner contour in \figref{fig:powermap}). We calculated the cross-correlation between the signal of every pixel and the reference signal, the maximum cross-correlation coefficient and lag time for each pixel were obtained by finding the argument maximum. These two values were plotted in  \figref{fig:powermap}{b-\ref{fig:powermap}c}. 

Similar analysis was done for the two minute bandpass (two to three minutes range), the Fourier power, cross-correlation coefficient, and lag time distributions  are shown in \figref{fig:powermap}{d-\ref{fig:powermap}f}.

\section{Structure and dynamics of the light bridge} 
\label{sec:results}

\subsection{\rev{Suppression of coronal loop formation}}

\figref{fig:fov_evol}{} presents the evolution of this sunspot and the associated active region  \AR with expanding coronal loops. An interesting feature is that on the  $10^\mathrm{th}$ and  $11^\mathrm{th}$ April 2019 this light bridge was  anchored at the north-west part of the penumbra (\figref{fig:fov_evol}{(a1-a2)}), no coronal loops were rooted at that part (\figref{fig:fov_evol}{(c1-c2)}). In contrast to this, after the bridge gradually detached from the penumbra, the coronal loops started to form again (\figref{fig:fov_evol}{(c4-c6)}). It appears that such a light bridge suppressed the formation of coronal loops above the part of umbra where it has rooted. We spotted that at the region where the light bridge was rooted, the strength and inclination angle of the magnetic field deviated from the other part of the penumbra (\figref{fig:fov}{(b-c)}). 	This part seems to have mixed polarity in contrast to the negative polarity at the umbra (see the part pointed by a blue arrow in \figref{fig:fov}{b}).

\subsection{Thermal and magnetic structure}
In this section and thereafter, we focus on the compact and segmented light bridge on 13 April 2019 (\figref{fig:fov_evol}{b4}). This structure could stand clear off the influence of complex dynamics, as its size was small and its magnetic structure might reveal the fundamental structure of a light bridge. 

\figref{fig:fov}{} illustrates the structure of the magnetic field and Doppler shift obtained by inversion of SOT/SP Stokes vector. We could see that, in the continuum intensity at $\lambda6301.5\unit{\AA}$ and the emission intensity of AIA $1700\unit{\AA}$, the elementary light bridge formed a compact bright island within umbra core (\figref{fig:fov}{a and \ref{fig:fov}d}). Its LOS magnetic field component $B_z$ was as weak as $-1400\unit{G}$, about half the strength of umbral field (about $-3000\unit{G}$). This variation was very sharp (about $1500\unit{G}$) and this was detected within a span of about $3\arcsec$ (or about $2,000\unit{km}$) across the bridge (\figref{fig:fov}{b}). Along the elongated spine of the bridge, the variation of LOS magnetic component was much smaller (about $500\unit{G}$). This fact also holds in the inclination angle of the magnetic field vector: along the spine, the inclination angle did not vary significantly, whereas across the spine we could discern that inclination angle at two flanks deviated by about $30\deg$ (or about $0.15\pi$, see \figref{fig:fov}{c}). The electric current density was strong at the northern end of the bridge, but no clear pattern was discernible.

In \figref{fig:fov}{f}, we observed red shift at the northern penumbra, its magnitude was about $1\unit{km\cdot s^{-1}}$. This means that fluid moved downwards to the sunspot. The Doppler shift was about $1\kmps$ at the outer edge of penumbra, this value was stronger than that at the penumbra-umbra interface. This is a clear pattern of Evershed flow \citep{Evershed1909}, which could be caused by overturning  magneto-convective motion guided by the magnetic field lines \citep{Rimmele2006,Siu-Tapia2018}. The umbra was also overwhelmed by red shift; at the bridge, there was two regions with flows either stronger or weaker than the background red shift. After removing the background red shift (about $0.6\kmps$), we observed paired patches of red and blue shifts on the Western and Eastern ends of this light bridge (\figref{fig:fov}{f}, zoomed view), their velocities were about $\pm0.3\kmps$.  

\subsection{Oscillations}

Within the umbral core, we could see that the emission intensity exhibited a persistent oscillatory signal at about two minutes, see \figref{fig:wavelet}. This oscillatory signal could be found within the whole umbra core, this is consistent with previous studies \citep{Yuan2014,Yuan2016lb,Jess2016}. The umbral oscillation power at two minutes became deplete at the light bridge. The cross-correlation of the two minute band was very low within the umbra (\figref{fig:powermap}{e  and \ref{fig:powermap}f}).  

At the light bridge, the emission intensity varied periodically at about four minutes (\figref{fig:wavelet}). This four minutes signal filled the light bridge (\figref{fig:powermap}{a}). This is consistent with earlier studies \citep{Yuan2014}. Two patches of strong oscillation power were measured at two ends of the bridge. This pattern overlapped with the paired regions with red and blue Doppler shift.  Within the bridge, the cross-correlation was very strong, the high-correlation region was found over the whole bridge (\figref{fig:powermap}{b}). The lag time of the four minute oscillation was close to zero within the bridge (\figref{fig:powermap}{c}).  

At the penumbra, similar four minutes oscillations were detected, the strong oscillation power was found to be distributed along the umbra-penumbra border. This is a commonly observed feature in sunspots \citep{Yuan2014,Yuan2016lb}.  This oscillatory signal was separate in space from the light bridge oscillation (\figref{fig:powermap}{a}). However, we notice that the four minutes oscillations at the penumbra and the light bridge had strong correlation (\figref{fig:powermap}{b}). As a contrast, the two minutes umbral oscillations had very weak correlation between any two locations (\figref{fig:powermap}{e}). The lag time at the penumbra could be either positive or negative, regions with same lag time tended to form patches in sizes of a few arc seconds (\figref{fig:powermap}{c}), it means that the four minute oscillation within the same patch could originate from a common oscillatory source from the underneath. 

\section{Conclusions and Discussions}
\label{sec:con}

In this letter, we used the joint observations of SDO/AIA and Hinode/SOT on \AR, and studied the long term evolution of a light bridge, its magnetic and thermal structures, and the oscillations of emission intensity. With a wealth of observables, we aim to infer the origin and structure of this light bridge. 

The light bridge appears to suppress the formation of coronal loops at the region where it is connected to the penumbra.  The region rooted by the light bridge have mixed polarity in contrast to the umbral field, this would means that the umbral field could connect to this polarity, rather than expand radially and form coronal loops. 

On the 13 April 2019, this bridge detached from the penumbra. A pair of red and blue shift regions were found at its two ends; whereas this whole structure was submerged within a global red shift. The paired patches of blue and red shift indicated upward and downward motion of plasma along the line of sight. The radiative MHD simulations of \citet{Rempel2011} and \citet{Toriumi2015a} suggest that magneto-convections could be guided by strong inclined magnetic field and plays a key role in triggering the dynamics above the light bridge.  The paired patches of blue and red Doppler shifts detected in this study implies that  this light bridge has a strong horizontal magnetic field component along its spine and directed the magneto-convection motions along its elongated part.  This effect has also been observed in cool loops \citep{Bethge2012}, emerging magnetic flux tube \citep{Requerey2017}, and sunspot umbra \citep{Kleint2013,Guglielmino2017,Guglielmino2019}. 

Within the umbra of \AR in this study, the emission intensity oscillate with a period of about two minute; however this oscillatory signal disappeared above the light bridge. As a contrast, this light bridge was a host of four minute oscillation in the emission intensity. Moreover, this periodicity is commonly found at the penumbra. So it appears that the four minute oscillation at the light bridge and that at the penumbra has a common origin, as they were highly correlated. If a light bridge is transported to the surface by magneto-convection \citep{Toriumi2015a}, its magnetic field could converge with that of the penumbra at a certain depth underneath the sunspot. That region could be the source of the four minutes periodicity. 

Here, we summarize the knowledge obtained within this letter. A pair of blue and red shift regions indicates that two ends of this light bridge are connected by a strong horizontal field. The paired upward and downward motions imply that this light bridge has \rev{two-legged} or undulate magnetic field.  If one end of this light bridge is rooted at the penumbra, it would suppress the formation of coronal loop at the anchor region. The four minute oscillation at the light bridge is highly correlated with that at the penumbra, it indicates that the light bridge and penumbra structure could be connected at a certain depth under the surface, they could have the same origin from a sub-surface magnetic field flux and become fragmented by magneto-convective motions \citep{Rempel2011,Toriumi2015a}.  With the flow pattern and oscillations, we could get an insight into the internal structure and formation of a light bridge, we should note that this light bridge only represents one kind, more cases are needed to probe the general structure of light bridges and investigate their influence on the dynamics and evolution of sunspots and pores. 
\acknowledgments
D.Y. is supported by the National Natural Science Foundation of China (NSFC, 11803005, 11911530690), Shenzhen Technology Project (JCYJ20180306172239618). S.F. is supported by the Joint Fund of NSFC (U1931107) and the Key Applied Basic Research program of Yunnan Province (2018FA035). Wavelet software was provided by C. Torrence and G. Compo, and is available at URL: \url{http://atoc.colorado.edu/research/wavelets/}. 

\bibliography{song2019lb}


\end{document}